\documentstyle[12pt,amssymb,epsf]{article}

\makeatletter
\@addtoreset{equation}{section}
\makeatother

\def\ben{\begin{equation}}
\def\een{\end{equation}}

\let\a=\alpha    
    
 \let\m=\mu \let\n=\nu

\let\C=\Chi

\def\nn{\nonumber} \def\bd{\begin{document}} \def\ed{\end{document}}
\def\ds{\documentstyle} \let\fr=\frac \let\bl=\bigl \let\br=\bigr
\let\Br=\Bigr \let\Bl=\Bigl
\let\bm=\bibitem
\let\na=\nabla
\let\pa=\partial \let\ov=\overline
\newcommand{\be}{\begin{equation}}
\newcommand{\ee}{\end{equation}}
\def\ba{\begin{array}}
\def\ea{\end{array}}
\def\ft#1#2{{\textstyle{{\scriptstyle #1}\over {\scriptstyle #2}}}}
\def\fft#1#2{{#1 \over #2}}
\def\del{\partial}
\def\vp{\varphi}
\def\sst#1{{\scriptscriptstyle #1}}
\def\oneone{\rlap 1\mkern4mu{\rm l}}
\def\td{\tilde}
\def\wtd{\widetilde}
\def\ie{\rm i.e.\ }
\def\dalemb#1#2{{\vbox{\hrule height .#2pt
        \hbox{\vrule width.#2pt height#1pt \kern#1pt
                \vrule width.#2pt}
        \hrule height.#2pt}}}
\def\square{\mathord{\dalemb{6.8}{7}\hbox{\hskip1pt}}}
\newcommand{\ho}[1]{$\, ^{#1}$}
\newcommand{\hoch}[1]{$\, ^{#1}$}
\newcommand{\bea}{\begin{eqnarray}}
\newcommand{\eea}{\end{eqnarray}}
\newcommand{\ra}{\rightarrow}
\newcommand{\lra}{\longrightarrow}
\newcommand{\Lra}{\Leftrightarrow}
\newcommand{\ap}{\alpha^\prime}
\newcommand{\bp}{\tilde \beta^\prime}
\newcommand{\tr}{{\rm tr} }
\newcommand{\Tr}{{\rm Tr} }
\def\0{{\sst{(0)}}}
\def\1{{\sst{(1)}}}
\def\2{{\sst{(2)}}}
\def\3{{\sst{(3)}}}
\def\4{{\sst{(4)}}}
\def\5{{\sst{(5)}}}
\def\6{{\sst{(6)}}}
\def\7{{\sst{(7)}}}
\def\8{{\sst{(8)}}}
\def\n{{\sst{(n)}}}
\def\cA{{{\cal A}}}
\def\cB{{{\cal B}}}
\def\cF{{{\cal F}}}
\def\tV{\widetilde V}
\def\tW{\widetilde W}
\def\tH{\widetilde H}
\def\tE{\widetilde E}
\def\tF{\widetilde F}
\def\tA{\widetilde A}
\def\im{{{\rm i}}}
\def\tY{{{\wtd Y}}}
\def\ep{{\epsilon}}
\def\vep{{\varepsilon}}
\def\R{\rlap{\rm I}\mkern3mu{\rm R}}
\def\bD{{{\bar D}}}

\def\R{\rlap{\rm I}\mkern3mu{\rm R}}
\def\bD{{{\bar D}}}
\def\R{{{\Bbb R}}}
\def\C{{{\Bbb C}}}
\def\H{{{\Bbb H}}}
\def\CP{{{\Bbb C}{\Bbb P}}}
\def\RP{{{\Bbb R}{\Bbb P}}}
\def\Z{{{\Bbb Z}}}
\def\bA{{{\Bbb A}}}
\def\bB{{{\Bbb B}}}
\def\bC{{{\Bbb C}}}
\def\bD{{{\Bbb D}}}
\def\bE{{{\Bbb E}}}
\def\bZ{{{\Bbb Z}}}
\def\Re{{{\frak{Re}}}}
\def\Im{{{\frak{Im}}}}
\def\cosec{{\,\hbox{cosec}\,}}
\def\Gm{{\Gamma_{\!\! -}}}
\def\Gp{{\Gamma_{\!\! +}}}
\def\stan{{standard }}
\def\nonstan{{supernumerary }}

\newcommand{\tamphys}{\it Center for Theoretical Physics,
Texas A\&M University, College Station, TX 77843}

\newcommand{\upenn}{\it Department of Physics and Astronomy,\\ University
of Pennsylvania, Philadelphia, PA 19104}

\newcommand{\brussels}{\it Physique Th\'eorique et Math\'ematique,
Universit\'e Libre de Bruxelles,\\ Campus Plaine C.P. 231, B-1050
Bruxelles, Belgium}

\newcommand{\auth}{S. Cucu\hoch{\ast1}, H. L\"u\hoch{\dagger2} and J.F.
V\'azquez-Poritz\hoch{\ddagger3}}

\begin{document}
\begin{flushright}

KUL-TF-03-04\ \ \ \ \
MIFP-03-02\ \ \ \ \
ULB-TH/03-03\\
March  2003\ \ \ \
\bf hep-th/0303211\\
\end{flushright}

\begin{center}
{\large {\bf A Supersymmetric and Smooth Compactification of\\
 M-theory to AdS$_5$}}

\vspace{10pt}
\auth

\vspace{10pt}
{\hoch{\ast}\it Instituut voor Theoretische Fysica, Katholieke
Universiteit Leuven,\\ Celestijnenlaan 200D B-3001 Leuven, Belgium}

\vspace{5pt}
{\hoch{\dagger}\it George P. and Cynthia W. Mitchell Institute for
Fundamental Physics,\\ Texas A\& M University, College Station, TX
77843-4242, USA}

\vspace{5pt}
{\hoch{\ddagger}\brussels}

\vspace{10pt}

\underline{ABSTRACT}
\end{center}

       We obtain smooth M-theory solutions whose geometry
is a warped product of AdS$_5$ and a compact internal space that
can be viewed as an $S^4$ bundle over $S^2$.  The bundle can be
trivial or twisted, depending on the even or odd values of the two
diagonal monopole charges.  The solution preserves ${\cal N}=2$
supersymmetry and is dual to an ${\cal N}=1$ $D=4$ superconformal
field theory, providing a concrete framework to study the
AdS$_5$/CFT$_4$ correspondence in M-theory. We construct analogous
embeddings of AdS$_4$, AdS$_3$ and AdS$_2$ in massive type IIA,
type IIB and M-theory, respectively. The internal spaces have
generalized holonomy and can be viewed as $S^n$ bundles over $S^2$
for $n=4$, 5 and 7. Surprisingly, the dimensions of spaces with
generalized holonomy includes $D=9$. We also obtain a large class
of solutions of AdS$\times H^2$.

{\vfill\leftline{}\vfill \vskip 10pt \footnoterule {\footnotesize
\hoch{1} Research supported in part by the Federal Office for
Scientific, Technical and Cultural Affairs \phantom{of the}
through the Interuniversity Attraction Pole P5/27 and the European
Community's Human \phantom{of the} Potential Programme under
contract HPRN-CT-2000-00131 Quantum Spacetime.

{\footnotesize \hoch{2}
Research supported in part by DOE grant DE-FG03-95ER40917.

{\footnotesize \hoch{3} Research supported in part by the Francqui
Foundation (Belgium), the Actions de Recherche \phantom{of the}
Concert{\'e}es of the Direction de la Recherche Scientifique -
Communaut\'e Francaise de Belgique, \phantom{of the} IISN-Belgium
(convention 4.4505.86) and by a ``Pole d'Attraction
Interuniversitaire." }} \vskip -12pt} } \pagebreak
\setcounter{page}{1}

\section{Introduction}

         AdS$_5$ spacetime arises naturally in type IIB supergravity,
which provides a non-trivial and relatively simple framework for
examining the holographic principle {\it via} the AdS$_5$/CFT$_4$
correspondence \cite{malda,gkp,wit}.  The embedding of AdS$_5$
spacetime in eleven-dimensional supergravity has also been studied
in the past.  A smooth but non-supersymmetric compactification of
eleven-dimensional supergravity to AdS$_5$ was obtained in
\cite{pn} where the internal space is a K\"ahler manifold. More
recently, an internal background of $\CP^2\times T^2$ was found in
\cite{dlp}. Although the compactification is not supersymmetric at
the level of supergravity, it was argued in \cite{dlp} that it is
fully supersymmetric at the level of M-theory, since it is T-dual
to the AdS$_5\times S^5$ of type IIB theory.  In the above two
examples, the AdS spacetime and internal manifold are direct
products without warp factors. Smooth but non-supersymmetric
M-theory solutions have been constructed in \cite{gauntlett1},
which are warped and twisted products of AdS$_5 \times S^2$ or
AdS$_5 \times H^2$ with a squashed four-sphere.

In \cite{ali,oz}, supersymmetric embeddings of AdS$_5$ in M-theory
were found as warped geometries with a compact internal metric.
This construction can be understood from that the fact that $S^5$
can be expressed as a foliation of $S^3$ and $S^1$.  One can then
T-dualize the AdS$_5\times S^5$ of type IIB theory on the $U(1)$
bundle of the $S^3$ and obtain a solution in M-theory \cite{clpv}.
However, there is a naked singularity in such a construction,
since the $U(1)$ circle of the $S^3$ can shrink to zero.
Supersymmetric and smooth embeddings of AdS$_5$ in M-theory were
obtained in \cite{mn1}.  The eleven-dimensional metric is a warped
product of AdS$_5$ with an internal metric that can be viewed as
an $S^4$ bundle over $H^2$, a hyperbolic 2-plane.  The
construction can give rise to both ${\cal N}=2$ and ${\cal N}=4$
supersymmetry.

       In this letter, we report a supersymmetric and smooth
compactification of M-theory to AdS$_5$, with the internal space being
an $S^4$ bundle over $S^2$.  The construction is only possible for
${\cal N}=2$ supersymmetry, and hence it gives rise to the minimum
AdS$_5$ gauged supergravity coupled to matter.  This solution provides
a supergravity dual to ${\cal N}=1$ $D=4$ superconformal field theory.
We also obtain supersymmetric and smooth compactifications of M-theory
to AdS$_2$ and type IIB to AdS$_3$. The internal space is an $S^p$
bundle over $S^2$, where $p=7$ and $5$, respectively. We also
construct a supersymmetric compactification of massive IIA to AdS$_4$,
which is singular.

\section{AdS$_5\times S^2$ in M-theory}

      We begin by considering the sector of $D=7$ gauged supergravity
with two diagonal $U(1)$ vector fields.  The relevant Lagrangian is
given by
\be
\hat e^{-1}{\cal L}_7 = \hat R - \ft12 (\del \phi_1)^2 -
\ft12 (\del \phi_2)^2 -
\hat V -\ft14 \sum_{i=1}^2 X_i^{-2}\, (\hat F_\2^i)^2\,,
\ee
where $X_i=e^{\ft12\vec a_i\cdot \vec \phi}$ with
\be
\vec a_1=(\sqrt2\,, \sqrt{\ft25})\,,\qquad
\vec a_2=(-\sqrt2\,, \sqrt{\ft25})\,.
\ee
The scalar potential $\hat V$ is given by \cite{tenauthors}
\be \hat V =g^2\, (-4X_1\, X_2 - 2 X_0\, X_1 - 2 X_0\, X_2 + \ft12
X_0^2) \,, \ee
where $X_0=(X_1\, X_2)^{-2}$.  The potential can be expressed in
terms of the superpotential
\be
\hat W=\fft{g}{\sqrt2}\, (X_0 + 2X_1 + 2 X_2)\,.
\ee
We now consider a 3-brane ansatz
\bea
ds^2&=&e^{2u}\, dx^\mu\, dx_\mu + e^{2v}\,
\lambda^{-2}\, d\Omega_2^2 +d\rho^2\,,\nn\\
F_\2^i&=&\ep\, m_i\, \lambda^{-2}\, \Omega_\2\,,\label{braneansatz}
\eea
where the constant $\ep$ takes the values 1, $-1$ and 0, if
$d\Omega_2^2$ is the metric for a unit $S^2$, hyperbolic $H^2$ or
2-torus $T^2$. $\Omega_\2$ is the corresponding volume form. The
system admits the following first-order equations
\bea
\fft{d\vec \phi}{d\rho} &=& \sqrt2\,\Big(-\fft{\epsilon}{2\sqrt2}\,
(m_1\, \vec a_1\, X_1^{-1} + m_2\, \vec a_2\, X_2^{-1})\,
e^{-2v} + \fft{d\hat W}{d\vec \phi}\Big)\,,\nn\\
\fft{dv}{d\rho} &=& -\fft{1}{5\sqrt2}\, \Big(2\sqrt2\,
\epsilon\,(m_1\, X_1^{-1} + m_2\, X_2^{-1})\, e^{-2v} +
\hat W\Big)\,,\nn\\
\fft{du}{d\rho} &=& \fft{1}{5\sqrt2}\, \Big(
\fft{\epsilon}{\sqrt2}\,(m_1\, X_1^{-1} + m_2\, X_2^{-1})\,
e^{-2v} - \hat W\Big)\,, \label{d7fo}
\eea
provided that the constraint
\be
\lambda^2=(m_1 + m_2)\, g\label{d7con}
\ee
is satisfied.  This set of first-order equations were derived for the
case of $H^2$ in \cite{mn1} by studying the Killing spinor equations
of $D=7$ gauged supergravity.  In \cite{clvs2}, a different method was
used to obtain them for $H^2$, $S^2$ and $T^2$.  The equations of
(\ref{d7fo}) were analysed in detail in \cite{clvs2}.  Here, we report
on only a subclass of solutions where $\vec \phi$ and $v$ are
constants.  In this case, for $\ep=0$, the solution is nothing but
AdS$_7$ written in Poincar\'e coordinates.

         For $\ep=\pm 1$, we find that the solution is given by
\bea
e^{\sqrt2 \phi_1} &=&
\fft{m_2 - m_1 \pm \sqrt{m_2^2 + m_1^2 - m_1\, m_2}}{m_2}\,,\qquad
e^{-\sqrt{\ft52}\, \phi_2} = \ft43\cosh(\phi_1/\sqrt2)\,,\nn\\
e^{-2v} &=& -\fft{\epsilon\, g\, e^{-\ft3{\sqrt{10}}\phi_2}}{
m_1\, e^{-\ft{\phi_1}{\sqrt2}} + m_2\, e^{\ft{\phi_1}{
\sqrt2}}}\,,\qquad u=-\ft12g\, e^{-\fft4{\sqrt{10}}\phi_2}\,
\rho\,.\label{d7sol} \eea
This solution is invariant under the simultaneous interchanges of
$m_1\leftrightarrow m_2$ and $\phi_1\leftrightarrow -\phi_1$.  The
reality conditions of the solution constrain the constants $m_i$ and
$g$, as well as the choice of $\pm$ in the solution.  Let us first
consider the case $\ep=-1$, corresponding to $d\Omega_2^2$ as the
metric of a unit (non-compact) hyperbolic 2-plane.  In this case, the
reality of the solution implies that $m_1\, m_2 \ge 0$.  This includes
the choice of $m_1=0$ (or $m_2=0$) and $m_1=m_2$, which were discussed
in \cite{mn1}.  The first case gives rise to ${\cal N}=4$
supersymmetry in $D=5$, whilst the second case gives rise to ${\cal
N}=2$ supersymmetry.

         We are particularly interested in a compact internal space.
Thus, we now turn to the choice of $\epsilon=+1$, corresponding to
$d\Omega_2^2$ as the metric of $S^2$.  In this case, the reality
conditions for (\ref{d7sol}) imply that $m_1\, m_2<0$.  The condition
(\ref{d7con}) implies further that $m_1\ne -m_2$.  Therefore, the
AdS$_5\times S^2$ solution can only have ${\cal N}=2$ supersymmetry,
but cannot arise from the pure $D=7$ minimal gauged supergravity.

        If we define a charge parameter $q=2m_1/(m_1 + m_2)$, then the
condition for having $S^2$ versus $H^2$ can be summarised as
\bea
q\in [0,2] &\Longrightarrow& H^2\,,\nn\\
q\in (-\infty,0)\,\, \hbox{or}\,\, (2,\infty) &\Longrightarrow&
S^2\,. \eea

      It is straightforward to lift this solution to $D=11$ by using
the ansatz obtained in \cite{tenauthors}.  Since the solutions for
general $m_i$ are rather complicated to present, we only consider a
representative example with $m_1=5g$ and $m_2=-3g$.  The M-theory
metric is given by
\bea
ds_{11}^2 &=& \Delta^{\ft13}\,\Big[ds_{\rm{AdS_5}}^2 +
\fft{1}{g^2\, c}\, \Big\{ \fft{1}{4c}\,
(d\theta^2 + \sin^2\theta\, d\varphi^2)\nn\\
&&+ \fft{1}{\Delta}\,\Big(\ft14\, d\mu_0^2 +
\ft15(d\mu_1^2 + \mu_1^2\, (d\phi_1 -\ft52 \cos\theta\, d\varphi)^2)\nn\\
&& + d\mu_2^2 + \mu_2^2\, (d\phi_2 + \ft32\cos\theta\, d\varphi)^2
\Big) \Big\}\Big]\,, \eea
where $c=10^{-2/5}$ and $\mu_i$ are spherical coordinates which
satisfy $\mu_0^2 + \mu_1^2 + \mu_2^2=1$.  The warp factor $\Delta$ is
given by
\be
\Delta=c\, (4\mu_0^2 + 5 \mu_1^2 + \mu_2^2)>0\,.
\ee
The AdS$_5$ metric is given by
\be
ds_{\rm{AdS_5}}^2 = e^{-\ft{2\rho}{R}}\, dx^\mu\, dx_\mu +
d\rho^2\,,
\ee
where the AdS radius is given by $R=\fft{1}{2c\,g}$.  The 4-form field
strength in $D=11$ can also be obtained using the reduction ansatz in
\cite{tenauthors}.  It is given by
\bea
{*F}_\4 &=& -(2\,g)^{-1}\,(8\mu_0^2 + 15\mu_1^2 + 7\mu_2^2)\,
\epsilon_\5\wedge \sin\theta\, d\theta\wedge d\varphi\\
&&+ g^{-1}\, \Big(\ft15 d(\mu_1^2)\wedge
(d\phi_1 -\ft52 \cos\theta\, d\varphi) - 3d(\mu_2^2)\wedge
(d\phi_2 + \ft32 \cos\theta\, d\varphi)\Big)\wedge \epsilon_\5\,.\nn
\eea
where $\epsilon_\5$ is the volume form for the AdS$_5$ metric.

     Thus, the internal space of the $D=11$ metric can be viewed as an
$S^4$ bundle over $S^2$, with two diagonal $U(1)$ bundles.  In
general, the internal metric can be labeled by $(q_1,
q_2)=(\fft{2m_1}{m_1+m_2}, \fft{2m_2}{m_1 + m2})$.  In the specific
example above, $(q_1,q_2)=(5,-3)$ and the solution is smooth
everywhere.  For general $(m_1,m_2)$, the metric does not have a
power-law singularity. However, it could have a conical orbifold
singularity, which is absent only if $(q_1,q_2)$ are integers.  Since
the $q_i$ satisfy the constraint $q_1 + q_2=2$, it follows that they
are either both even or both odd integers. In the even case, the
bundle is topologically trivial, whilst it is twisted for the odd
case.

\section{AdS$_4\times S^2$ in massive type IIA}

          The scalar potential in gauged supergravity with two $U(1)$
isometries was obtained in \cite{gubser}.  From this, we deduce that
the relevant Lagrangian involving the two $U(1)$ vector fields is
given by
\be
\hat e^{-1}{\cal L}_6 = \hat R - \ft12 (\del \phi_1)^2 -
\ft12 (\del \phi_2)^2 -
\hat V -\ft14 \sum_{i=1}^2 X_i^{-2}\, (\hat F_\2^i)^2\,,
\ee
where $X_i=e^{\ft12\vec a_i\cdot \vec \phi}$ with
\be
\vec a_1=(\sqrt2\,, \fft1{\sqrt2})\,,\qquad
\vec a_2=(-\sqrt2\,, \fft1{\sqrt2})\,.
\ee
The scalar potential is given by
\be
\hat V=\ft49 g^2\, (X_0^2 - 9 X_1\, X_2 - 6 X_0\, X_1 - 6 X_0\, X_2)\,,
\ee
where $X_0=(X_1\,X_2)^{-3/2}$.  As in the previous case, the scalar
potential can be expressed in terms of the superpotential
\be
\hat W=\fft{g}{\sqrt2}\,(\ft43 X_0 + 2 X_1 + 2 X_2)\,.
\ee

      We consider a membrane solution of the type given by
(\ref{braneansatz}).  The system admits the following first-order
equations
\bea
\fft{d\vec \phi}{d\rho} &=& \sqrt2\,\Big(-\fft{\epsilon}{2\sqrt2}\,
(m_1\, \vec a_1\, X_1^{-1} + m_2\, \vec a_2\, X_2^{-1})\,
e^{-2v} + \fft{d\hat W}{d\vec \phi}\Big)\,,\nn\\
\fft{dv}{d\rho} &=& -\fft{1}{4\sqrt2}\, \Big(\fft3{\sqrt2}\,
\epsilon\,(m_1\, X_1^{-1} + m_2\, X_2^{-1})\, e^{-2v} +
\hat W\Big)\,,\nn\\
\fft{du}{d\rho} &=& \fft{1}{4\sqrt2}\, \Big(
\fft{\epsilon}{\sqrt2}\,(m_1\, X_1^{-1} + m_2\, X_2^{-1})\,
e^{-2v} - \hat W\Big)\,, \label{6}
\eea
provided that the constraint $\lambda^2=(m_1+m_2)g$ is satisfied. The
solutions were analysed in detail in \cite{clvs2}.  Here, we shall
only consider the subset of solutions with constant scalars.  For
$\epsilon=0$, one just reproduces the AdS$_6$ metric in Poincar\'e
coordinates.  For $\epsilon=\pm 1$, we have
\bea {\rm e}^{\sqrt{2}\phi_1}&=&\frac{3}{2}\, \frac{m_2-m_1\pm
\sqrt{(m_2-m_1)^2+\frac{4}{9}m_1\, m_2 }}{m_2}\,,\qquad {\rm
e}^{-\sqrt{2}\phi_2}=\frac{3}{2} {\rm
cosh}(\frac{\phi_1}{\sqrt{2}})\,,\nn\\
{\rm e}^{-2v}&=& -\frac{4\epsilon\,g\, {\rm
e}^{-\frac{\phi_2}{\sqrt{2}}}}{m_1\, {\rm
e}^{-\frac{\phi_1}{\sqrt{2}}}+m_2\, {\rm
e}^{\frac{\phi_1}{\sqrt{2}}}}\,,\qquad u=-\frac{g}{3}{\rm
e}^{-\frac{3}{\sqrt{8}}\phi_2}\rho\,. \label{ads4omega2}
\eea
As in the $D=7$ result, we can define a charge parameter
$q=\fft{2m_1}{m_1 + m_2}$.  We have $H^2$ or $S^2$ depending on the
following condition:
\bea
q\in [0,2] &\Longrightarrow& H^2\,,\nn\\
q\in (-\infty,0)\,\, \hbox{or}\,\, (2,\infty) &\Longrightarrow&
S^2\,. \label{d6rule} \eea
When $q=0$ or $q=2$, the system has ${\cal N}=4$ supersymmetry.
Otherwise, we have ${\cal N}=2$ supersymmetry.

         Using the reduction ansatz in \cite{gubser,massive}, it is
straightforward to lift the solution back to $D=10$, giving rise to a
solution of massive type IIA supergravity.  The metric is given by
\bea ds_{10}^2 &=& \mu_0^{\ft1{12}}\, X_0^{\ft18}\, (X_1\,
X_2)^{\ft14} \,\Delta^{\ft38}\, \Big[ds_6^2 + g^{-2}\,
\Delta^{-1}\,
\Big(X_0^{-1}\, d\mu_0^2\\
&&+X_1^{-1}\, (d\mu_1^2 + \mu_1^2\,(d\varphi_1 + g\,A_\1^2)^2) +
X_2^{-1}\, (d\mu_2^2 + \mu_2^2\,(d\varphi_2 + g\,A_\2^1)^2)\Big)\Big]
\,,\nn
\eea
where $\Delta=\sum_{\alpha=0}^2 X_\a\,\mu_\alpha^2>0$ and $\m_0^2 +
\mu_1^2 + \mu_2^2=1$.  Thus, the $D=10$ metric is a warped product of
AdS$_4$ with an internal six-metric, which is an $S^4$ bundle over
$S^2$ or $H^2$, depending on the charge parameter $p$, according to
the rule (\ref{d6rule}).

As an example of a supersymmetric, though singular,
compactification of AdS$_4$ from massive IIA, we can take $m_1=7g$
and $m_2=-5g$, and a choice of negative sign in
(\ref{ads4omega2}). This gives $X_0=6c$, $X_1=7c$ and $X_2=c$,
where $c=6^{-1/4}\, 7^{-3/8}$. Also, $A_\1^1=-\frac{7}{2g}{\rm
cos}\, \theta\, d\varphi$ and $A_\1^2=\frac{5}{2g}{\rm cos}\,
\theta\, d\varphi$, and the radius of AdS$_4$ is given by
$R=1/(2c\, g)$.

\section{AdS$_3\times S^2$ in type IIB}

Let us now consider the $D=5$ minimal gauged supergravity
coupled to two vector multiplets. The Lagrangian is given by
\be e^{-1}{\cal L}_5=\hat R - \ft12 (\del \phi_1)^2 -\ft12 (\del
\phi_2)^2 - \fft14\sum_{i=1}^3 X_i^{-2} (\hat F_\2^i)^2 - \hat V +
e^{-1}\, \ft14 \epsilon^{\mu\nu\rho\sigma\lambda}\, \hat
F^1_{\mu\nu}\,\hat F^2_{\rho\sigma}\, \hat A^3_\lambda\,, \ee
where $X_i=e^{\ft12\vec a_i\cdot\vec \phi}$ with
\be
\vec a_1=(\sqrt2\,, \fft{2}{\sqrt6})\,, \qquad
\vec a_2=(-\sqrt2\,,\fft2{\sqrt6})\,,\qquad \vec
a_3=(0,- \,\fft4{\sqrt6})\,.
\ee
The scalar potential is given by
\be
\hat V=-4g^2 \sum_{i=1}^3
         X_i^{-1}\,.\label{stdscalarpot}
\ee
The scalar potential $\hat V$ can also be expressed in terms of
the superpotential
\be
\hat W=\sqrt2\, g\, \sum_i X_i\,.
\ee
We find that the string solution of the type given by (\ref{braneansatz})
admits the following first-order equations
\bea \fft{d\vec \phi}{d\rho} &=&
\sqrt2\,\Big(-\fft{\epsilon}{2\sqrt2}\, (m_1\, \vec a_1\, X_1^{-1}
+ m_2\, \vec a_2\, X_2^{-1} + m_3\, \vec a_3\, X_3^{-1})\,e^{-2v}
+
\fft{d\hat W}{d\vec \phi}\Big)\,,\nn\\
\fft{dv}{d\rho} &=& -\fft{1}{3\sqrt2}\,
\Big(\sqrt2\, \epsilon\,(m_1\, X_1^{-1} +
m_2\, X_2^{-1} + m_3\, X_3^{-1})\, e^{-2v} + \hat W\Big)\,,\nn\\
\fft{du}{d\rho} &=& \fft{1}{3\sqrt2}\, \Big(
\fft{\epsilon}{\sqrt2}\,(m_1\, X_1^{-1} + m_2\, X_2^{-1} + m_3\,
X_3^{-1})\, e^{-2v} - \hat W\Big)\,, \label{5}
\eea
provided that the constraint $\lambda^2=(m_1+m_2+m_3)g$ is
satisfied. The general solution for this system was analysed in
\cite{clvs2}. Here, we consider only the solutions with constant
scalar fields.  For $\epsilon=0$, the solution is AdS$_5$ in
Poincar\'e coordinates.  For $\ep=\pm 1$, fixed-point solutions exist
only for non-vanishing $m_i$. The solution is given by
\bea
e^{\sqrt2\, \phi_1} &=& \fft{m_1}{m_2}\, \Big(\fft{m_3+m_2-m_1}{m_3 -
m_2 + m1}\Big)\,,\qquad
e^{\sqrt6\, \phi_2} = \fft{m_1\,m_2\,(m_3^2 - (m_1-m_2)^2)}{
m_3^2\,(m_1 + m_2 - m_3)^2}\,,\nn\\
e^{-2v}&=&-\epsilon\,g\,\Big(\fft{(m_1 + m_2 - m_3)(m_3^2 -
(m_1-m_2)^2)}{m_1^2\,m_2^2\,m_3^2}\Big)^{\ft13}\,,\nn\\
u&=& -g\, e^{\fft{\phi_2}{\sqrt6}}\, \Big(\cosh(\phi_1/\sqrt2) +
\ft12 e^{-\sqrt{\ft32}\,\phi_2}\Big)\, \rho\,.
\eea
The reality condition of the solution implies that when three vectors
with the magnitudes $|m_i|$ can form a triangle, $d\Omega_2^2$ should
be the $H^2$ metric. On the other hand, when they cannot form a
triangle, the metric should be that of $S^2$.\footnote{AdS$_3\times
S^2$ solutions were also recently found in \cite{sabra} in a different
construction.}  If any of the $m_i$ vanish, there is no fixed-point
solution, except when one vanishes with the remaining two equal
\cite{clvs2}.

        Using the reduction ansatz obtained in \cite{tenauthors}, we
can easily lift the solution back to $D=10$.  Since the solution
with general $m_i$ is complicated to present, we consider a
simpler case with $m_2=m_1$.  The $D=10$ type IIB metric is
\bea
ds_{10}^2&=& \sqrt{\Delta}\Big\{ ds_{\rm{AdS_3}}^2 +\epsilon\,g^{-2}
(\fft{m_1}{m_3-2m_1})^{1/3}\, (\ft12 q_1\, d\Omega_2^2 + d\theta^2)\nn\\
&&+ g^{-2}\,\Delta^{-1}\,\Big[ c^{-1/3}\, \cos^2\theta\,\Big
(d\psi^2 + \sin^2\psi\, (d\varphi_1 + \ft12 q_1\, A_\1)^2\nn\\
&& +\cos^2\psi\, (d\varphi_2 + \ft12 q_1\, A_\1)^2\Big) +
c^{2/3}\, \sin^2\theta\, (d\varphi_3 + \ft12 q_3\,
A_\1)^2\Big]\Big\}\,, \eea
where
\bea
&&c=\Big|\fft{m_1}{2m_1-m_3}\Big|\,,\qquad \Delta=c^{1/3}\, \cos^2\theta
+ c^{-2/3} \sin^2\theta>0\,,\qquad dA_\1=\Omega_\2\,,\nn\\
&&ds_{\rm{AdS_3}}^2 = e^{-\fft{2\rho}{R}}\,, (-dt^2 + dx^2) +
d\rho^2\,, \qquad R=\Big|\fft{2m_1}{g\,(4m_1-m_3)\, c^{1/3}}\Big|
\eea
We have introduced the charge parameters $q_i=2 m_i/(m_1 + m_2 +
m_3)$, and hence they satisfy the constraint $q_1 + q_2 + q_3=2$.  In
the above solution, if $|m_3|< 2|m_1|$, we should have $\epsilon=-1$,
corresponding to $H^2$; if $|m_3|> 2|m_1|$, we should have
$\epsilon=1$, corresponding to $S^2$.  In general, the internal metric
is an $S^5$ bundle over $S^2$ or $H^2$, depending the values of the
$q_i$ according to the above rules.

\section{AdS$_2\times S^2$ in M-theory}

      Let us now consider the $U(1)^4$ gauged $N=2$ supergravity in
four dimensions. The Lagrangian is given by
\be e^{-1}{\cal L}_4=\hat R - \ft12 (\del \phi_1)^2 -\ft12 (\del
\phi_2)^2 -\ft12 (\del \phi_3)^2 -\fft14\sum_{i=1}^4 X_i^{-2}
(\hat F_\2^i)^2 - \hat V \,, \ee
where $X_i=e^{\ft12\vec a_i\cdot\vec \phi}$ with
\be
\vec a_1=(1, 1, 1),\,\quad \vec a_2=(1, -1, -1),\,\quad \vec a_3=(-1, 1,
-1),\,\quad \vec a_4=(-1, -1, 1)\,.
\ee
The scalar potential is given by
\be \hat V=-4g^2 \sum_{i<j} X_i\, X_j\,, \label{d4V} \ee
which can also be expressed in terms of the superpotential
\be \hat W=\sqrt2\, g\, \sum_{i=1}^4 X_i\,. \ee
The magnetic black hole solution of the type given by
(\ref{braneansatz}) admits the following first-order equations
\bea \fft{d\vec \phi}{d\rho} &=&
\sqrt2\,\Big(-\fft{\epsilon}{2\sqrt2}\, \sum_{i=1}^4 m_i\, \vec
a_i\, X_i^{-1}\, \,e^{-2v} +
\fft{d\hat W}{d\vec \phi}\Big)\,,\nn\\
\fft{dv}{d\rho} &=& -\fft{1}{2\sqrt2}\,
\Big(\fft{\epsilon}{\sqrt{2}}\,
\sum_{i=1}^4 m_i\, X_i^{-1}\, e^{-2v} + \hat W\Big)\,,\nn\\
\fft{du}{d\rho} &=& \fft{1}{2\sqrt2}\, \Big(
\fft{\epsilon}{\sqrt2}\, \sum_{i=1}^4 m_i\, X_i^{-1}\, e^{-2v} -
\hat W\Big)\,, \label{4} \eea
provided that the constraint $\lambda^2=g\sum_{i=1}^4 m_i$ is
satisfied. The general solution for this system was analysed in
\cite{clvs2}. Here, we consider only solutions with constant scalar
fields.  For $\epsilon=0$, the solution is AdS$_4$ in Poincar\'e
coordinates.  For $\epsilon=\pm 1$, we have not obtained the general
solution for abitrary $m_i$, although we have found many examples of
specific solutions.  We do find a class of special solutions by
setting $m_2=m_3=m_4$.  This enables us to consistently set
$\phi_1=\phi_2=\phi_3\equiv \phi$. For this truncation, fixed-point
solutions for $\epsilon = \pm 1$ are given by
\bea
{\rm e}^{2 \phi} &=& \frac{3m_2-m_1\pm
\sqrt{(m_1-m_2)(m_1-9m_2)}}{2m_2}\,,\qquad
{\rm e}^{-2v}= 4g\, \epsilon\, \frac{{\rm sinh}\, \phi}{m_1\, {\rm
e}^{-2\phi}-m_2}\,,\nn\\
u&=& -\ft12g\,\Big( \frac{2(m_1\, {\rm
e}^{-\frac{3}{2}\phi}+3m_2\, {\rm e}^{\frac{1}{2}\phi})}{m_2-m_1\,
{\rm e}^{-2\phi}}\, {\rm sinh}\, \phi+{\rm
e}^{\frac{3}{2}\phi}+3{\rm e}^{-\frac{1}{2}\phi}\Big)\, \rho\,.
\eea
The reality condition of the solution implies that for $\ep=-1$,
corresponding to $H^2$, we must have either $m_2 >0$ and $0<m_1\le
m_2$ or $m_2<0$ and $m_2\le m_1 < -3m_2$.  For $\ep=1$,
corresponding to $S^2$, we must have $m_2\le 0$ and $m_1>-3m_2$.
The AdS$_2 \times H^2$ has also been found in \cite{gauntlett2}.

        Using the reduction ansatz obtained in \cite{tenauthors}, we
can easily lift the solution back to $D=11$ with the metric
\bea
ds_{11}^2 &=& \Delta^{2/3}\Big\{ ds_{\rm{AdS_2}}^2 +\frac{{\rm
e}^{2v}}{(m_1+3m_2)g}d\Omega_2^2\nn\\
& &+\frac{1}{g^2\, \Delta} \Big[ {\rm
e}^{-\frac{3}{2}\phi}\Big( d\mu_1^2+\mu_1^2(d\phi_1+\frac{\epsilon\,
m_1}{m_1+3m_2} A_\1)^2 \Big)\nn\\
& &+{\rm e}^{\frac{1}{2}\phi} \sum_{i=2}^4 \Big( d\mu_i^2+\mu_i^2
(d\phi_i +\frac{\epsilon\, m_2}{m_1+3m_2} A_\1)^2 \Big) \Big]
\Big\}\,, \eea
where
\bea
\Delta &=& ({\rm e}^{\frac{3}{2}\phi}-{\rm
e}^{-\frac{1}{2}\phi})\mu_1^2+{\rm e}^{-\frac{1}{2}\phi}>0\,,
\quad dA_\1=\Omega_\2\,,\quad
ds_{\rm{AdS_2}}^2 = -e^{-\fft{2\rho}{R}}\, dt^2 + d\rho^2\,,
\nn\\
R &=& \frac{2}{g} \Big[ \frac{2(m_1\, {\rm e}^{-\frac{3}{2}\phi}+3m_2\,
{\rm e}^{\frac{1}{2}\phi})}{m_2-m_1\, {\rm e}^{-2\phi}}\, {\rm sinh}\,
\phi+{\rm e}^{\frac{3}{2}\phi}+3{\rm e}^{-\frac{1}{2}\phi}\Big]^{-1}\,.
\eea
In general, the internal metric is an $S^7$ bundle over $S^2$ or
$H^2$, depending the values of the $m_i$.

\section{Conclusions}

      We have obtained a large class of supersymmetric embeddings of
AdS spacetime in M-theory, as well as type IIB and massive type
IIA theories.  The internal spaces can be viewed as $S^n$ bundles
over $S^2$ or $H^2$.  Similar solutions have been discussed in
\cite{gauntlett3,gauntlett4,gauntlett5}.  In particular, we have
found a smooth embedding of AdS$_5$ in M-theory, with a compact
internal space of an $S^4$ bundle over $S^2$.  The bundle can be
trivial or twisted, depending on the two diagonal monopole
charges.  The solution preserves ${\cal N}=2$ supersymmetry; it is
a supergravity dual to an ${\cal N}=1$, $D=4$ superconformal field
theory on the boundary of AdS$_5$. This provides a concrete
framework to study AdS$_5$/CFT$_4$ from the point of view of
M-theory.

        The internal spaces of these embeddings may
be regarded as concrete realisations of spaces with generalized
holonomy groups advocated in \cite{duffliu}, since they are not
Ricci flat and involve a form field.  An especially interesting
example is the $S^7$ bundle over $S^2$ or $H^2$, which is
nine-dimensional. While nine-dimensional Ricci-flat manifolds do
not have an irreducible special holonomy group, our aforementioned
solutions are explicit examples of nine-dimensional spaces which
have generalized special holonomy.

      The embeddings of AdS spacetimes in M-theory and string theories
discussed in this paper all involve warp factors.  We expect that
there are many further examples of such solutions.  It is of interest
to explore them, both from the AdS/CFT perspective as well as for a
more concrete understanding and classification of spaces with
generalized special holonomy.

\section*{ACKNOWLEDGMENT}

        We are grateful to Gary Gibbons, Chris Pope and Ergin Sezgin
for useful discussions.

\end{document}